\documentclass[12pt,preprint]{aastex}

\usepackage{natbib}
\usepackage{graphicx}
\newcommand{\eqb}{\begin{equation}}
\newcommand{\eqe}{\end{equation}}

\newcommand{\Ai}{{\rm Ai}}
\newcommand{\Bi}{{\rm Bi}}
\begin{document}

\title{Curvature-drift instability fails to generate pulsar radio emission}

\author{Alexander Kaganovich, Yuri Lyubarsky}
\affil{Physics Department, Ben-Gurion University, P.O.B. 653, Beer-Sheva 84105, Israel}

\begin{abstract}
The curvature drift instability has long been considered as a viable mechanism for pulsar radio emission. We reconsidered this mechanism by finding an explicit solution describing propagation of short-wave electro-magnetic waves in a plasma flow along curved magnetic field lines. We show that even though the waves could be amplified, the amplification factor remains very close to unity therefore this mechanism is unable to generate high brightness temperature emission from initial weak fluctuations.
\end{abstract}


\maketitle

\section{Introduction}

It is generally believed that the coherent pulsar radio emission is generated by some sort of instability
in the relativistic electron-positron plasma ejected from the pulsar along the open field line tube.
One of the widely recognized candidates is the curvature instability associated with plasma motion along the curved field lines. \citet{goldreich_keeley71} found an electro-magnetic instability of a set of charged particles moving on a circular ring. On the contrary \citet{blandford75} demonstrated that maser action is impossible in the plasma flow
along infinitely strong magnetic field unless points of inflection and torsion are present in the field geometry. The controversy was resolved by \citet{asseo_pellat_sol83} and \citet{larroche_pellat87} who showed that the instability develops only if the flow is sharply bounded; the necessary conditions are unlikely to be met in pulsar magnetospheres. \citet{zheleznyakov_shaposhnikov79} noticed that the wave amplification is possible even in broad, smooth outflows if the inertial drift of particles is taken into account. This curvature-drift
instability was further investigated by
\citet{kazbegi_mach_melik91,luo_melrose92,lyut_bland_mach99,lyut_mach_bland99,shapakidze03,osmanov09}.

The curvature-drift instability belongs to the class of resonant instabilities, i.e. it develops when and if particles move in phase with the wave 
so that the power of the electric force applied to the particles, $\bf E\cdot v$, does not oscillate.
The necessary condition for the resonance is that the wave phase velocity is less than the speed of light.
The transverse wave in a strongly magnetized plasma does become subluminal; in the electron-positron plasma
moving along the magnetic field  with the Lorentz factor $\gamma_p$, the phase velocity of the wave propagating at a small angle to the magnetic filed is (e.g. \citet{kazbegi_mach_melik91})
 \eqb
\frac{\omega}{kc}=1-\delta;\qquad \delta=\frac{\omega^2_p}{4\omega_c^2\gamma_p^3};
 \label{phase_velocity}\eqe
where
  \eqb
 \omega_p=\sqrt{\frac{4\pi e^2n}{m}};\qquad \omega_c=\frac{eB}{mc}.
  \eqe
Here $e$ and $m$ are the electron charge and mass, correspondingly, $n$ the pair number density.  The above simple dispersion law is obtained if the wave frequency in the plasma frame is below the cyclotron frequency, $\omega\ll\omega_c/\gamma_p$, and if one neglects the energy spread of the plasma particles.

Conventional wisdom is that the electron-positron plasma is generated near the pulsar polar cap by a highly relativistic primary particle beam accelerated by rotationally induced electric field. Such a two-component flow ideally suits to the curvature drift instability because a single component could not experience a significant drift, which requires a large Lorentz factor, and at the same time to affect sufficiently the wave dispersion properties.
In the two-component flow, the dense, relatively low energy secondary plasma makes the wave subluminal according to Eq. (\ref{phase_velocity}) whereas the energetic primary beam has a drift velocity sufficient to resonantly excite the wave.

 Because the magnetic field in the pulsar magnetosphere is very large, the factor $\delta$ is very small therefore 
the wave velocity remains very close to the speed of light.
Since the beam is highly relativistic,
the resonance condition, $\omega=\mathbf{k\cdot v}_b$, could be fulfilled only if the wave propagates at a very small angle to the beam. However, the radiation propagates along nearly straight rays (the refraction index is close to unity) whereas the particle beam follows the curved magnetic field lines therefore for any ray, the resonance condition could be fulfilled only within a small zone where the ray nearly grazes the field line (see Fig. 1). 

\begin{figure}
\includegraphics[width=15 cm,scale=0.5]{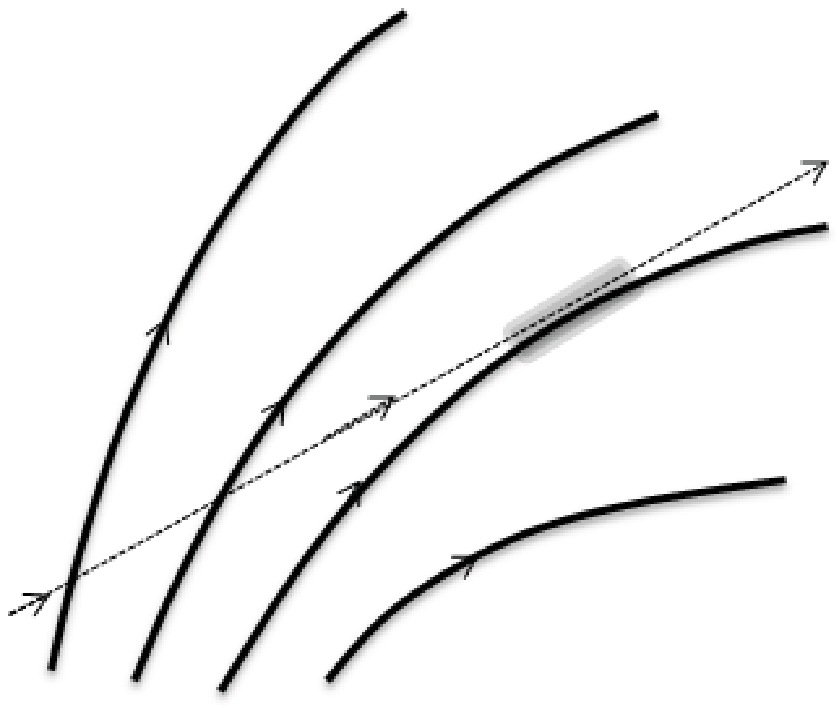}
\caption{Geometry of the problem. Radiation propagates along nearly straight rays (dotted arrow) across the plasma flowing along the curved magnetic field lines (thick arrows). The resonance amplification is possible only if the angle between the wave propagation direction and the particle velocity is small. This condition could be fulfilled only within a small region (shaded) where the ray nearly grazes the magnetic field. Therefore the amplification factor of the wave is limited.}
\end{figure}

Thus the wave is resonantly amplified only at a short length scale. This is in spite of the fact that the size of the unstable region could be large, i.e. one can find unstable waves at any point within the region. However, any such a wave rapidly leaves the resonant range of the angles when propagates in the curved magnetic field. It is the resonant growth \textsc{length that limits} the amplification factor of the wave.
The curvature instability could be considered as a viable mechanism for pulsar radio emission if the amplification factor is exponentially large; only in this case small electro-magnetic fluctuations in the flow could give rise to the observed high brightness temperature emission. We will show in this paper
that this never happens; the amplification factor in any case remains very close to unity.

 The paper is organized as follows. In the next section, we outline the basic parameters of the pulsar magnetosphere. In sect. 3, we present preliminary, qualitative analysis of the wave amplification by curvature drift instability. In sect. 4, we derive the wave equations and introduce the coordinate system appropriate for analyzing the wave amplification.In sect. 5, we find response of the medium, i.e. calculate electric currents excited in the main plasma flow and in the high energy beam by the wave. The amplification factor of the wave is found in sect. 6.

\section{Plasma flows in pulsar magnetosphere}

 Let us first shortly outline the basic parameters of the pulsar plasma.
Better conditions for the curvature drift instability are achieved in the external part of the pulsar magnetosphere where the magnetic field is not too high. Assuming the dipole field, one finds the cyclotron frequency at the distance $D$ from the star as
 \eqb
\omega_c=1.8\cdot 10^{10}\frac{B_{*12}}{d^3_3}\,\rm Hz;
 \eqe
where $B_*$ is the surface field, $d=D/r_*$,  $r_*$ the star radius and the usual notation $A=10^xA_x$ is used.

The particle density in the primary beam is of the order of Goldreich-Julian particle density,
$n_b=B/(eP)$, where $P$ is the pulsar period. Therefore the plasma frequency of the beam may be estimated as
 \eqb
\omega_b^2\equiv\frac{4\pi e^2n_b}{m}=\frac{4\pi\omega_c}P.
 \label{omega_b}\eqe
Note that both the particle density and the magnetic field in the open field line tube vary inversely proportionally to the tube cross section therefore the expression (\ref{omega_b}) remains valid in any point.

 The Lorentz factor of the primary beam is very high, $\gamma_b\sim 10^6$,
therefore they experience inertial drift when moving along the curved magnetic field lines. The drift velocity is directed along the binormal to the field line;
 \eqb
\mathbf{u}_d=c\frac{\mathbf{F\times B}}{eB^2};
 \eqe
where $F=\gamma_b mc^2/R_c$ is the centrifugal force, $R_c$ the curvature radius of the field line. Within the open field line tube, the curvature radius could be estimated as \eqb
R_c=\sqrt{DR_L}=2\cdot 10^9\sqrt{d_3P}\,\rm cm;
\label{radius}\eqe
where
$R_L=P/2\pi$ is the light cylinder radius. Now one gets an estimate
 \eqb
|u_d|/c=8\cdot 10^{-4}\gamma_{b6}d_3^{2.5}B_{*12}^{-1}P^{-1/2}.
 \eqe
One sees that the drift velocity is very small. Therefore the optimal conditions for the curvature-drift instability are achieved in the region of the pulsar light cylinder where the drift velocity is maximal.

The secondary electron-positron plasma
is generated with a moderate Lorentz factor $\gamma_p\sim 10\div 100$. The plasma density is generally assumed to be large as compared with the Goldreich-Julian density.  Introducing the multiplicity factor, $\lambda$, defined as the ratio of the plasma to the Goldreich-Julian density, one can write
\eqb
\omega_p^2=\lambda\omega_b=\frac{4\pi\lambda\omega_c}P.
\eqe
According to the available polar cap models, $\lambda$
varies from a few to a few thousands \citep{hibsch_arons01}. Observations of pulsar wind nebulae give evidence for much larger multiplicities, $\lambda> {\rm few}\times 10^5$ \citep{dejager07}.

 Now one can estimate the parameter $\delta$, which defines, according to Eq. (\ref{phase_velocity}), the dispersion properties of the electro-magnetic waves in pulsar magnetospheres:
 \eqb
\delta=1.7\cdot 10^{-6}\frac{\lambda_4 d^3_3}{\gamma_p^3B_{*12}P}.
 \eqe
One sees that the phase velocity of the waves is very close to the speed of light at any reasonable pulsar conditions.

\section{Preliminary analysis of the problem}

\begin{figure}
\includegraphics[width=9 cm,scale=1.2]{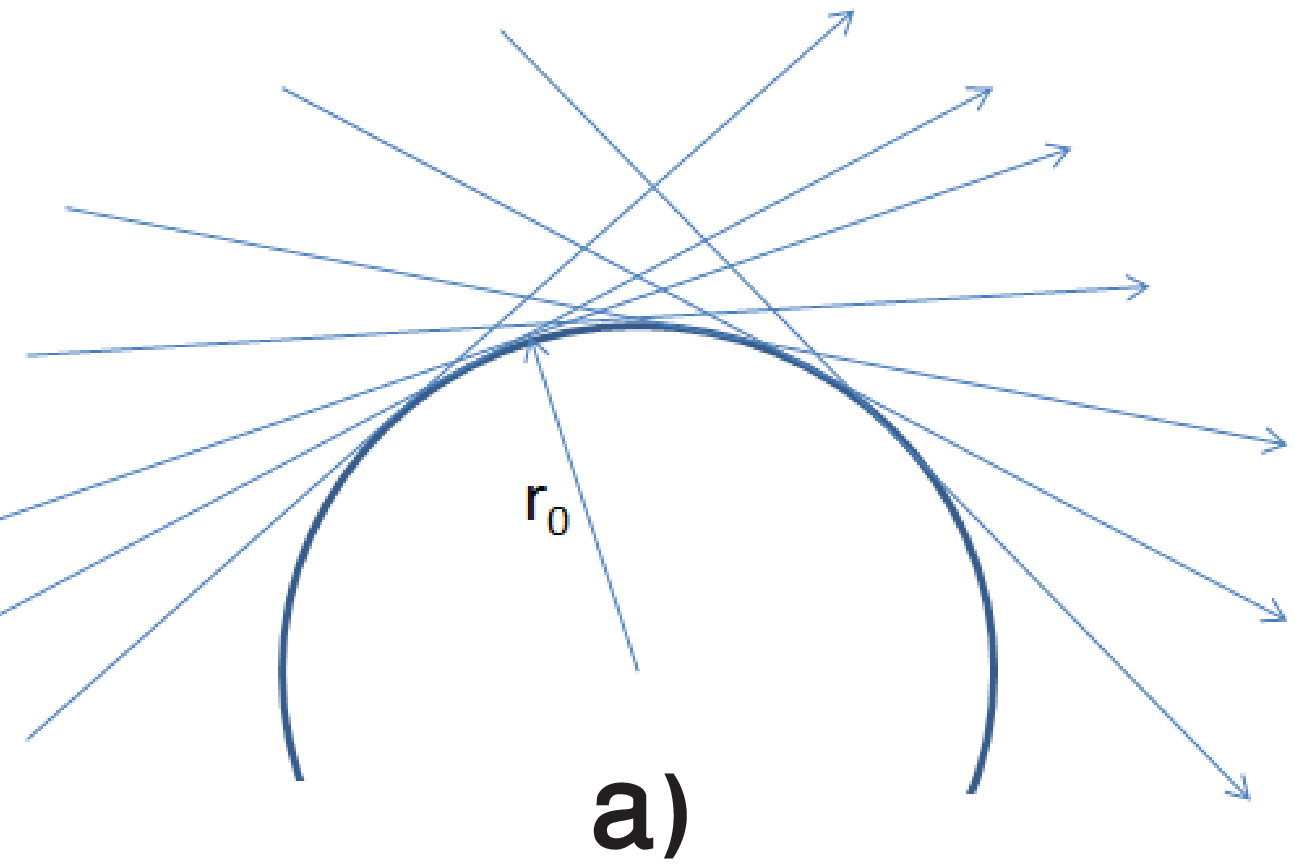}
\includegraphics[width=9 cm,scale=1.2]{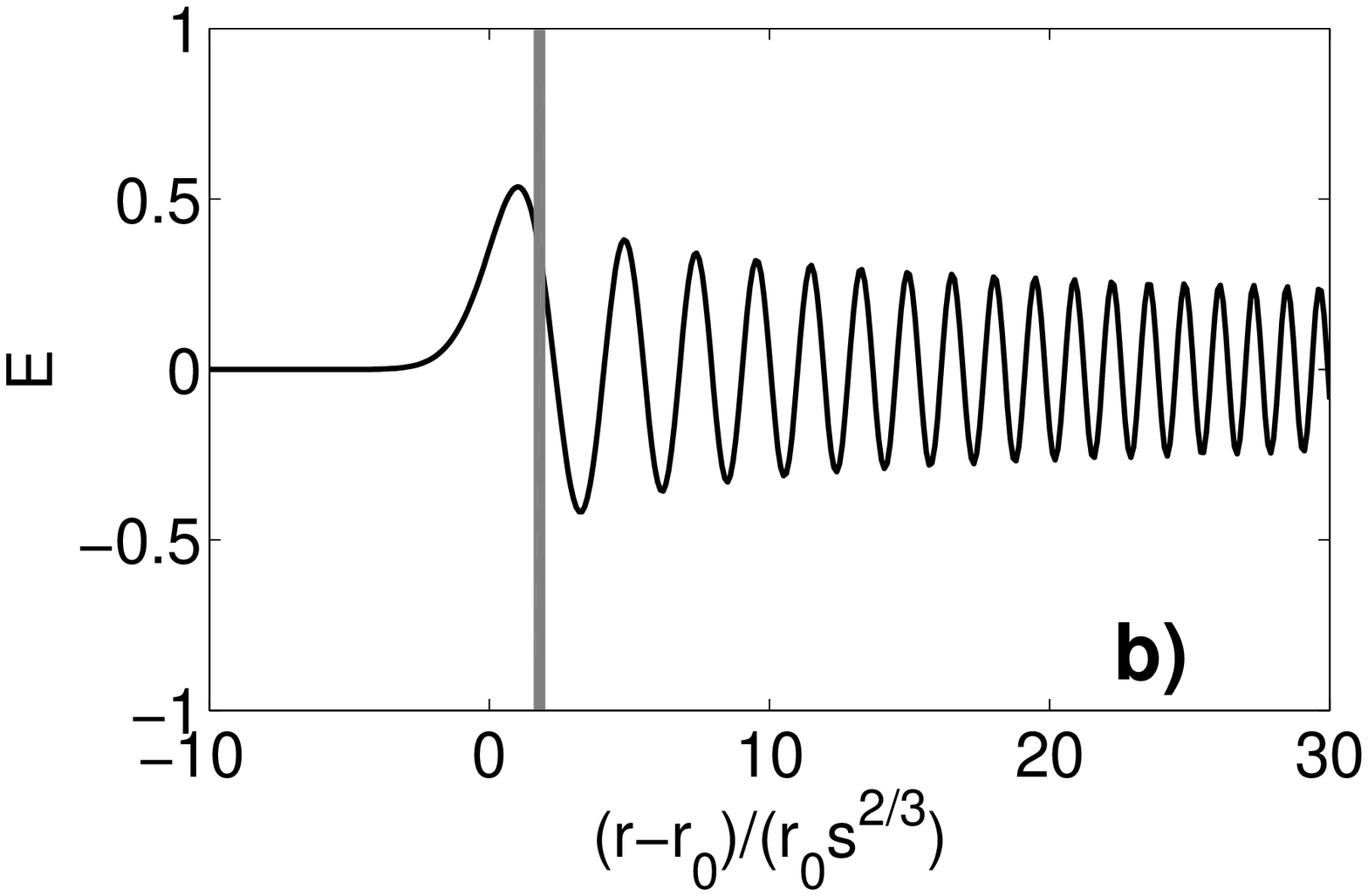}
\caption{The structure of the cylindrical wave. a) In the WKB approximation, the wave is comprised from a rays (thin arrows) coming from all directions with the same impact parameter, $r_0$. At $r=r_0$ they form a caustic (thick circle). b) The radial structure of the wave in the caustic zone. The vertical stripe shows the position of the resonance layer.}
\end{figure}

 We are going to study the resonant amplification of radio waves when they propagate across the pulsar plasma. As it was already discussed in the Introduction, the wave could be resonantly amplified only if it propagates close to the direction of the particle beam.
The particles move along the curved magnetic field lines and also could experience a weak inertial drift along the binormal to the field line. Therefore the amplifying wave should be directed at a small angle to the magnetic field and polarized predominantly along the direction of the drift velocity. Since the radiation propagates along nearly straight rays whereas the beam particles follow the curved magnetic field lines,
the amplification could occur only in a small zone (Fig. 1). The magnetic field lines within this zone could be considered as arches of concentric circles along which both the plasma and the beam circulate. This configuration approximates locally real plasma motion along curved magnetic field lines in the pulsar magnetosphere. Therefore one can conveniently use the cylindrical coordinates $(r,\phi, z)$ with
the $\phi$ direction along the background magnetic field.

In this configuration, one can expand the wave in cylindrical modes,
 \eqb
 \mathbf{E}(\mathbf{r},t)= \mathbf{E}_{\omega,s,k_z}(r)\exp(-i\omega t+ik_zz+is\phi);
 \label{cylinder_wave}\eqe
 thus reducing the problem to an ordinary differential equation for  $\mathbf{E}_{\omega,s,k_z}(r)$.
The wavelength is very small as compared with the characteristic inhomogeneity scale,
\eqb
\omega r/c=4\cdot 10^8f_{\rm GHz}\sqrt{d_3P};
\eqe
where $f_{\rm GHz}$ is the wave frequency in GHz; we used the estimate (\ref{radius}) for the curvature radius. Therefore one can try to use the WKB approximation.

In this case, the solution should take a form of locally plane waves satisfying the dispersion equation  (\ref{phase_velocity}), i.e. there should be
 \eqb
\mathbf{E}(\mathbf{r},t)\propto A_{\rm out}e^{i\int k_rdr} +A_{\rm in}e^{-i\int k_rdr};
 \label{WKB}\eqe
where the radial wave vector, $k_r$, should be found from the dispersion equation (1) as
 \eqb
k_r^2=\frac{\omega^2}{c^2(1-\delta)^2}-\frac{s^2}{r^2}-k_z^2.
 \label{dispersion}\eqe
The first term in Eq. (\ref{WKB}) represents rays propagating away from the axis. At any point, the direction of such a ray is along the wave vector $(k_r,k_{\phi}=s/r,k_z)$. Taking into account Eq. (\ref{dispersion}), one finds the impact parameter of the ray,
 \eqb
 r_0=\frac{cs}{\sqrt{\left(\frac{\omega}{1-\delta}\right)^2-c^2k_z^2}}.
  \eqe
We are interested in rays propagating together with the beam therefore $k_z/\omega\sim u_d\ll 1$. Then one can expand
 \eqb
 r_0=\frac{cs}{\omega}\left(1-\delta+\frac 12\frac{c^2k_z^2}{\omega^2}\right).
  \label{impact}\eqe
The second term in (\ref{WKB}) represents rays propagating towards the axis with the impact parameter (\ref{impact}). Therefore the cylindrical wave could be considered as a set of rays with the same impact parameter  (Fig. 2a).
These rays converge to the axis till the minimal distance $r_0$ and then go to the infinity.
Two rays, ingoing and outgoing, pass through any point at $r>r_0$.  At $r\approx r_0$, a caustic is formed where the WKB approximation is violated. The structure of the solution should resemble that of the Airy function (Fig. 2b): the maximum near $r_0$, oscillations at $r>r_0$ and the exponential decay (the evanescent wave) at $r<r_0$. The width of the caustic is very small, $\Delta r=r_0^{1/3}(c/\omega)^{2/3}\ll r_0$. The oscillating part approaches the WKB solution (\ref{WKB}) at $r-r_0\gg\Delta r$.

In the $r-\phi$ plane, the lines of constant phase rotate with a constant angular velocity $\omega/s$. At $r>r_0$, they form unwinding spirals (see Fig.2 in \citet{lyut_mach_bland99}) approaching eventually the lines $\int k_rdr+s\phi/r=\it const$. Normals to these lines form rays plotted in Fig. 2a.
The azimuthal component of the wave phase velocity $v_{{\rm ph},\phi}\equiv\omega/k_{\phi}=r\omega/s$ varies with the radius.
Therefore the resonant interaction with particles could occur only in a narrow layer where the particle azimuthal velocity is close to  $r\omega/cs$.

More exactly, the resonant condition is $\omega=\mathbf{k\cdot v}$. The particle azimuthal velocity is $v_{\phi}=\sqrt{c^2-u_d^2-1/(2\gamma_b^2)}\approx c[1-u_d^2/(2c^2)]$ therefore the resonance condition is written as
 \eqb
\omega=k_zu_d+(cs/r)[1-u^2_d/(2c^2)].
\eqe
Resolving this equation with respect to $r$, one finds the radius of the resonance layer
 \eqb
r_{\rm res}=\frac{cs}{\omega}\left[1+\frac{u_d}c\left(\frac{ck_z}{\omega}-\frac{u_d}{2c}\right)\right].
 \label{resWKB}\eqe
The condition that the wave reaches the resonance radius, $r_0< r_{\rm res}$, can now be written as  \citep{lyut_bland_mach99}
 \eqb
\delta>\frac 12\left(\frac{k_zc}{\omega}-\frac{u_d}c\right)^2.
 \label{cond_WKB}\eqe
The right-hand side of this expression goes to zero if the ray is inclined to the magnetic field line plane at the same angle as the particle beam (more accurate calculation shows  that the right-hand side goes not to zero but to $\gamma_b^{-2}$, which could be neglected) therefore at first glance, this condition does not impose any restrictions on $\delta$. However, we will see that the amplification also ceases in this case therefore some discrepancy between the directions of the ray and the beam is necessary. The reason is that the resonance interaction occurs via the component of the wave electric field parallel to the particle velocity, $\mathbf{E\cdot v}$. The transverse wave should be inclined to the beam in order for this component be non-zero. Then Eq. (\ref{cond_WKB}) implies that the curvature drift mechanism could operate only if the magnetospheric plasma is dense enough.

Since both $\delta$ and $u_d/c$ are very small at any reasonable pulsar conditions, the ray should be inclined to the plane of the magnetic field line by a small angle. In this case, the resonance and the caustic radii are close to each other. Therefore in order to find the amplification factor of the wave, we could solve the wave equations in a relatively narrow range of radii containing both $r_0$ and $r_{\rm res}$. At the external boundary of this region, the solution should be matched with the WKB solution (\ref{WKB}). At the internal boundary, the solution should be matched with the evanescent wave. Then the amplification factor is found as the ratio of amplitudes of the outgoing and ingoing waves, $a=|A_{\rm in}|/|A_{\rm out}|$. With such a setup, we have in fact to solve the reflection problem for the radial wave $\mathbf{E}_{\omega,s,k_z}(r)$. Therefore the curvature instability may be thought of as a partial case of the general effect, namely, overreflection of waves from a rotating dielectric body \citep{zeldovich72}.

\section{Wave equations}
The propagation of the wave is described by Maxwell's equations
 \begin{eqnarray}
\nabla\cdot\mathbf{E}=4\pi\rho; & \nabla\times\mathbf{B}=4\pi\mathbf{j}+
\frac{\partial\mathbf{E}}{\partial t}; &\\
  \nabla\cdot\mathbf{E}=0; &
\nabla\times\mathbf{E}=-\frac{\partial\mathbf{B}}{\partial t};&
 \end{eqnarray}
where currents and charges are excited in the plasma by the wave. From here on we take the speed of light to be unity.
One can conveniently study the curvature instability by expanding the wave in cylindrical modes (\ref{cylinder_wave}).
Eliminating $\mathbf{B}$, one
writes the equations for the amplitude of a cylindrical wave as
 \begin{eqnarray}
 \frac{d^2 E_{\phi}}{d r^2}+\frac 1r \frac{d E_{\phi}}{d r}
+\left(\omega^2-k^2_z-\frac{1}{r^2}\right)E_{\phi}
-i\frac s{r}\frac{d }{d r}\frac{E_r}r+\frac{k_zs}rE_z
& =& -4\pi \omega ij_{\phi};\label{phi}\\
\frac{d^2 E_z}{d r^2}+\frac 1r \frac{d E_z}{d r}
+\left(\omega^2-\frac{s^2}{r^2}\right)E_z
-i\frac{k_z}{r}\frac{d }{d r}rE_r+\frac{k_zs}rE_{\phi}
& =&  -4\pi \omega i j_z;\label{z} \\
\left(\omega^2-k^2_z-\frac{s^2}{r^2}\right)E_r
-is\frac{d}{d r}rE_{\phi}-ik_z\frac{dE_z}{dr}
& =&  -4\pi\omega ij_r;\label{r}
 \end{eqnarray}

We are going to solve these equations close to the caustic zone where the wave propagates nearly along the $\phi$ direction (which coincides with the direction of the background magnetic field). 
We will see that the wave remains nearly transverse therefore one can conveniently introduce the variables
 \eqb
E_+=\frac s{\omega r}E_{\phi}+\frac{k_z}{\omega}E_z;\quad E_-=\frac s{\omega r}E_z
-\frac{k_z}{\omega}E_{\phi}
 \label{E+-}\eqe
such that in the region of interest, $E_+$ becomes the longitudinal component of the field whereas
$E_-$ and $E_r$ the transverse components. Combining Eqs. (\ref{phi}) and (\ref{z}), one
gets the equation for $E_-$
 \begin{eqnarray}
 \frac{d^2 E_{-}}{dr^2}+\frac 1r\left(1+\frac{s^2}{\omega^2r^2}\right) \frac{dE_{-}}{dr}
+\left[\omega^2-k^2_z-\frac{s^2}{r^2}\left(1+\frac 1{s^2}+\frac 1{\omega^2r^2}\right)\right]E_{-} \nonumber\\
+\frac{k_zs}{\omega^2r^2}\frac{dE_+}{dr}-\frac{2sk_z}{\omega r^2}iE_r=  -4\pi i\omega j_{-};
 \label{E-2}\end{eqnarray}
where 
 \eqb
j_-=\frac s{\omega r}j_z-\frac{k_z}{\omega}j_{\phi}.
 \eqe
We complement this equation by Eq. 
(\ref{r}) for $E_r$ and by the Gauss law
 \eqb
\frac 1r \frac{drE_r}{d r}+i\omega E_+=4\pi\rho.
 \label{gauss}\eqe

Since $\omega r\approx s\gg 1$, we can solve these equations in the short wave approximation (e.g. \citet{nayfe}). The resonance occurs close to the caustic zone therefore one can conveniently use the caustic coordinates. The plasma refraction index is very close to unity therefore the characteristic width of the caustic zone is the same as in the vacuum case, $\Delta r\sim r/s^{2/3}$; the reflection point is just shifted towards a smaller radius. Therefore
let us define the new radial dimensionless coordinate, $x$, as
 \eqb
r=r_0\left(1+\frac x{s^{2/3}}\right);
 \label{x}\eqe
where $r_0$ is the WKB impact parameter (\ref{impact}), which is in fact the caustic radius.
Transforming to the new variable and retaining only the leading order in $s^{-1/3}\approx(\omega r_0)^{-1/3}$ terms, we can reduce the equations (\ref{E-2}), (\ref{r}) and (\ref{gauss}) to the form
\begin{eqnarray}
 E_-''+2(x-s^{2/3}\delta)E_- &=& -\frac{4\pi s^{2/3}}{\omega} ij_-;\label{eqE-}\\
 is^{1/3} E'_+ -2(x-s^{2/3}\delta)E_r&=& \frac{4\pi s^{2/3}}{\omega}ij_r; \label{eqE+}\\
 E_r'+is^{1/3} E_+ &=& \frac{4\pi s^{1/3}}{\omega}\rho;\label{eqEr}
 \end{eqnarray}
where prime denotes differentiation with respect to $x$.

In order to write the equations in the closed form, we have to find the charge and current densities excited by the wave in the plasma and in the beam. Here we consider the simplest case
when both the plasma and the beam are cold so that they are described only by their densities and
velocities. Then the current and the charge densities are presented as
 \eqb
\rho=\sum_{\alpha}q_{\alpha}n_{\alpha};\quad \mathbf{j}=\sum_{\alpha}q_{\alpha}n_{\alpha}\mathbf{v}_{\alpha}; \eqe
where the summation is over the particle species (electrons and positrons in the plasma and, say,
positrons in the beam). Now we have to find perturbation of the particle motion in the
electro-magnetic field of the wave.

\section{The response of the medium}
In this section, we find the current and charge density induced in the medium by the wave.
With this purpose, we have to solve the equations of particle motion in the field of the wave:
 \eqb
\frac{\partial \gamma\mathbf{v}}{\partial t}+(\mathbf{v}\cdot\nabla)\gamma\mathbf{v}=\frac qm(\mathbf{E}+\mathbf{v\times B}_0 +\mathbf{v\times B}).
 \label{motion}\eqe
Here $q=\pm e$ is the particle charge; the background field, $\mathbf{B}_0$,
is assumed to have only $\phi$ component. The oscillatory motion of the particles in the electro-magnetic field of the wave implies perturbation of the particle density according to the continuity equation
 \eqb
 \frac{\partial n}{\partial t}+(\nabla\cdot n\mathbf{v})=0.
 \label{continuity}\eqe
 Without the wave, the particles move
in the $\phi$ direction along the background magnetic field and also experience the inertial drift in
the $z$ direction.
In order to find the response of the medium to the wave, one has to linearize Eqs. (\ref{motion}) and (\ref{continuity})
with respect to small perturbations of the velocity, $\mathbf{v}=\mathbf{v}^{(0)}+\mathbf{v}^{(1)}$
and of the density $n=n^{(0)}+n^{(1)}$. The response is expressed via the solution to the linearized equations as
 \eqb
 \rho=\sum_{\alpha}q_{\alpha}n_{\alpha}^{(1)};\quad \mathbf{j}=\sum_{\alpha}q_{\alpha}(n_{\alpha}^{(1)}\mathbf{v}_{\alpha}^{(0)}+
 n_{\alpha}^{(0)}\mathbf{v}_{\alpha}^{(1)}).
\eqe
Even though the procedure is straightforward, it leads to rather
cumbersome expressions therefore
we consider response of the plasma and of the beam separately exploiting from the
beginning appropriate approximations.

\subsection{Response of the beam}

The Lorentz factor of the beam is very large, $\gamma_b\sim 10^6$, therefore when moving along the curved magnetic field line, it experiences the inertial drift in the $z$ direction
 \eqb
v^{(0)}_{z,\rm beam}\equiv u_d=-\frac{\gamma_b V^2}{r\omega_c}.
\label{u_d}\eqe
The unperturbed velocity in the $\phi$ direction may be presented as
 \eqb
v^{(0)}_{\phi,\rm beam}\equiv V=\sqrt{1-u_d^2-\gamma_b^{-2}}.
 \label{V}\eqe
In this subsection, only velocities and density of the beam appear therefore we drop the index "beam" in $\mathbf{v}^{(1)}$ and $n^{(1)}$.
Linearizing the continuity equation (\ref{continuity}) one gets
 \eqb
n^{(1)}=\frac{n_b}{\Omega}\left(k_zv^{(1)}_z+\frac srv^{(1)}_{\phi}-
i\frac{\partial v^{(1)}_r}{\partial r}\right);
 \label{cont}\eqe
where
 \eqb
\Omega=\omega-V\frac sr-k_zu_d+i0;
 \eqe
and we take into account that according to the Landau prescription, a small
imaginary part should be added to the frequency.

The linearized equations of motion are written as
 \begin{eqnarray}
i\Omega v_{r}^{(1)}+\frac{\omega_c}{\gamma_b}v_z^{(1)}+\frac{2V}rv_{\phi}^{(1)}&=&
\frac e{m\gamma_b}(VB_z-E_r-u_dB_{\phi}); \label{vr}\\
i\Omega v_{\phi}^{(1)}-\frac Vrv_{r}^{(1)}&=&\frac e{m\gamma_b}\left[
Vu_dE_z-\left(u^2_d+\frac 1{\gamma_b^2}\right)E_{\phi}+u_dB_r\right]; \label{vphi}\\
i\Omega v_{z}^{(1)}-\frac{\omega_c}{\gamma_b}v_r^{(1)}&=&
-\frac e{m\gamma_b}[(1-u^2_d)E_z+VB_r-u_dVE_{\phi}]. \label{vz}
 \end{eqnarray}
The terms with $V/r$ in the left-hand side of Eqs. (\ref{vr}) and (\ref{vphi}) arise
from the centrifugal and Coriolis forces, correspondingly. These terms were neglected
in all previous works on the curvature instability. We will see that they are important
near the resonance where $\Omega$ goes to zero.

The density of the beam is extremely small therefore contribution of the beam to the current and charge densities of the medium is significant only near the resonance layer where the wave moves together
with the beam so that where $\Omega$ goes to zero. The resonant radius is defined from the condition $\Omega=0$,  which is nothing more than the condition $\omega=\mathbf{k\cdot v}_b$ already discussed in Sect. 3. Therefore we can use Eq. (\ref{resWKB}) for $r_{\rm res}$.
Expressing $r_{\rm res}$ via the radial coordinate $x$ defined by Eq. (\ref{x})
and making use of Eq. (\ref{impact}) for $r_0$,
one gets the position of the resonance layer in the dimensionless variable
 \eqb
x_{\rm res}=\delta s^{2/3}
-\frac 12s^{2/3}(1-\kappa)^2u_d^2.
 \label{xres}\eqe
 Taking into account that the ray is directed nearly along the particle beam, one can expect $k_z/\omega\sim u_d$ therefore we introduced the parameter
 \eqb\kappa=\frac{k_z}{u_d\omega}.\eqe
The function $\Omega$ can now be written via the dimensionless variable as
 \eqb
\Omega=\frac{\omega}{s^{2/3}}(x-x_{\rm res}+i0).
 \eqe

When solving Eqs. (\ref{vr}-\ref{vz}) for $v_{\phi}^{(1)}$ and $v_z^{(1)}$,  we retain only resonant terms (those proportional either to $\Omega^{-1}$ or $\partial/\partial r$).
Expressing $B$ via $E$ with the aid of Maxwell's equations, one finds
 \begin{eqnarray}
v_{\phi}^{(1)}&=&-i\frac{e}{m\gamma_bV}
\left\{\frac 1{\Omega}\left[\frac 1{\gamma_b^2}-\kappa(1-\kappa)u_d^4\right]-
\frac{u_d^2}{\omega V^2}r\frac{d}{dr}\right\}\left(VE_{\phi}+u_dE_z\right);\label{phi_res}\\
v_z^{(1)}&=&-i\frac{eu_d}{m\gamma_bV^2}\left\{\frac{2}{\Omega}\left[\frac 1{\gamma_b^2}-\kappa(1-\kappa)u_d^4\right]+\frac{u_d}{\omega }
r\frac{d}{dr}\right\}\left(VE_{\phi}+u_dE_z\right).\label{z_res}
 \end{eqnarray}
The component $v_r^{(1)}$ has no resonant terms however, it should be taken into account in the expression (\ref{cont}) for
$n^{(1)}$ because being differentiated, it could contribute to the resonance interaction provided the field amplitudes are singular.
Dropping the terms proportional to $\Omega$ one finds
 \eqb
 v_r^{(1)}=-\frac{eu_d}{m\omega_c}(1-\kappa)\left(VE_{\phi}+u_dE_z\right).
 \label{r_res}\eqe

One sees that the resonance interaction is determined by the
projection of the wave electric field onto the direction of the beam. This projection is expressed via the
"longitudinal" and "transverse" components of the electric field (\ref{E+-}) as
 \eqb
 VE_{\phi}+u_dE_z=\left(\frac{sV}{\omega r}+\frac{u_dk_z}{\omega}\right)E_++
 \left(\frac{u_ds}{\omega r}-\frac{Vk_z}{\omega}\right)E_-\approx E_++(1-\kappa)u_dE_-.
 \eqe

Substitution of Eqs. (\ref{phi_res}), (\ref{z_res}) and (\ref{r_res}) into Eq.
(\ref{cont}) yields the beam contribution to the charge density in the form
 \eqb
4\pi i\rho^{\rm beam}=4\pi ie n^{(1)}=-\frac{\omega^2_bs^{4/3}u_d^4\kappa(1-\kappa)}{\omega\gamma_b(x-x_{\rm res}+i0)^2}
\left(1+(x-x_{\rm res})\frac{d}{dx}\right)\left[E_++(1-\kappa)u_dE_-\right].
 \label{rho_res}\eqe
Here we made transformation to the dimensionless coordinate $x$ and also dropped the term $1/\gamma^2_b$ as compared with $u_d^4$ in the square brackets in Eqs. (\ref{phi_res})
and (\ref{z_res}).
 One sees that the density perturbation is of the second order in $\Omega^{-1}\propto (x-x_{\rm res})^{-1}$
whereas the velocities are only of the first order. Therefore the beam contribution
to the current could be written as
\eqb
\mathbf{j}^{\rm beam}=\rho^{\rm beam}\mathbf{v}^{(0)}_{\rm beam},
\eqe
which yields
 \eqb
j_{-}^{\rm beam}=\rho^{\rm beam}\left(\frac{k_z}{\omega}V-\frac{s}{\omega r_0}u_d\right)\approx(1-\kappa)u_d\rho^{\rm beam}.
 \label{j_res}\eqe
Note that if we were neglected the $V/r$ terms (arising from the Coriolis and centrifugal forces) in the linearized equations of motion (\ref{vr}) and (\ref{vphi}), we would come to the expression similar to (\ref{rho_res}) but with the coefficient proportional to $u_d^2$ instead of $u_d^4$. Therefore taking into account these terms is crucially important.

\subsection{Response of the plasma}
The Lorentz factor of the plasma particles, $\gamma_p=(1-v_p^2)^{-1/2}$, is not very large so that one can neglect the inertial forces. Then the unperturbed velocity is in the $\phi$ direction,
$v^{(0)}_z=0$; $v^{(0)}_{\phi}\equiv v_p$. In this subsection, we drop the index "plasma" in $\mathbf{v}^{(1)}$ and $n^{(1)}$. The linearized equations of motion are  now written as
 \begin{eqnarray}
i\left(\omega-\frac{sv_p}r\right)v_{r}^{(1)}+\frac{\omega_c}{\gamma_p}v_z^{(1)}&=&
\frac q{m\gamma_p}(v_pB_z-E_r); \\
i\left(\omega-\frac{sv_p}r\right)v_{\phi}^{(1)}&=&-\frac q{m\gamma_p^3}E_{\phi};\\
i\left(\omega-\frac{sv_p}r\right) v_{z}^{(1)}-\frac{\omega_c}{\gamma_p}v_r^{(1)}&=&
-\frac q{m\gamma_p}(E_z+v_pB_r).
 \end{eqnarray}
Since the plasma is electrically neutral in the sense that the densities and unperturbed velocities of electrons and positrons are equal, the plasma response is determined only by terms with odd powers of $q$ in the perturbed velocity. These terms are easily found as
 \begin{eqnarray}
v_r^{(1)}&=&-i\frac{e\gamma_p(\omega-v_ps/r)}{m\omega_c^2\omega}\left[(\omega-v_ps/r)E_r-
iv_p\frac{dE_{\phi}}{dr}\right]+\{{\rm terms}\propto e^2\}; \\
v_{\phi}^{(1)}&=&i\frac{e}{m\gamma^3_p(\omega-v_ps/r)}E_{\phi};\\
v_z^{(1)}&=&-i\frac{e\gamma_p(\omega-v_ps/r)}{m\omega_c^2\omega}\left[(\omega-v_ps/r)E_{z}+
v_pk_zE_{\phi}\right]+\{{\rm terms}\propto e^2\}.
 \end{eqnarray}
Here we used the strong field approximation,
 \eqb
 \omega_c\gg\gamma_p(\omega-sv_p/r)\approx\omega/2\gamma_p.
\label{highfield}\eqe

The charge and current densities are written as
 \eqb
\rho^{\rm plasma}=2en^{(1)}=\frac{2en_p}{\omega-v_ps/r}\left(\frac srv_{\phi}^{(1)}+k_zv_z^{(1)}
-\frac ir\frac{d rv_r^{(1)}}{dr}\right);
 \eqe
 \eqb
j_{\phi}^{\rm plasma}=v_p\rho+2en_pv_{\phi}^{(1)};\qquad
j_{z}^{\rm plasma}=2en_pv_{z}^{(1)};\qquad j_{r}^{\rm plasma}=2en_pv_{r}^{(1)}.
 \eqe
Here $n_p$ is the pair number density (the number density of electrons or positrons); the factor 2 takes into account that contributions of electrons and positrons are equal.
Note that the plasma current and charge densities are small as $\delta\sim (\omega_p/\omega_c)^2$ (see Eq. (\ref{phase_velocity})). Then the "longitudinal" component of the electric field, $E_+$, is also small as $\delta$
(or as $s^{-1/3}$, see Eq.(\ref{eqEr})) therefore
one can neglect the contribution of this component to the plasma current and charge
densities. Then 
one gets
 \begin{eqnarray}
4\pi ij_{-}^{\rm plasma} &=& 2\left\{\frac{\omega_p^2\gamma_p}{\omega^3\omega^2_c}
\left[\frac{\omega s}r-v_p\left(k_z^2+\frac{s^2}{r^2}\right)\right]\left(\frac{\omega s}r-
\frac{v_ps^2}{r^2}-k_z^2\right)-\right. \nonumber \\
&& \left. \frac{\omega^2_p k_z^2}{\gamma_p^3(\omega-v_ps/r)}\right\}E_-
 -2\frac{\omega_p^2k_z^2}{\omega^3\omega^2_c}\frac{d^2 E_-}{d r^2}
+\frac{\omega_p^2k_z}{\gamma_p\omega^2_c}\frac{dE_r}{d r}.
 \end{eqnarray}
Here we took into account that $dE_-/d r\gg E_-/r$.

These expressions could be significantly simplified taking into account that we are interested in the region close to the resonance layer defined by Eq. (17) so that we can substitute $r_{\rm res}$ instead of $r$.
Then one can write, e.g.,
 \eqb
\omega-\frac{sv_p}{r}=\omega\left(1-v_p\frac{1-\kappa u_d^2}V\right)\approx\frac{\omega}{2\gamma^2_p}.
 \eqe
In the last equality, we take into account that $\gamma_p|u_d|\ll 1$. Making use of this approximation and
taking also into account the condition (\ref{highfield}), one finds
 \eqb
4\pi ij_{-}^{\rm plasma}=\omega^2_p\left(\frac{\omega}{2\omega_c^2\gamma_p^3}-\frac{8\gamma_pk_z^2}{\omega^3}\right)E_-
+\frac{\omega_p^2k_z}{\omega_c^2}\left(i\frac{\omega}{\gamma_p s^{1/3}}E'_r
-\frac{2k_z}{s^{2/3}}E''_-\right).
 \eqe
The term with $E'_r$ can be neglected because one can show with the aid of Eq. (\ref{eqEr}) that this term is small as compared with the term with $E_-$ (small as $E_+/E_-$
or as $\omega E_-/\rho\sim\delta$). The term with $E''_-$ could also be neglected because when substituting the expression
for $j_{-}^{\rm plasma}$ into the right-hand side of Eq. (\ref{eqE-}), one sees that this term is small as compared with the corresponding term in the left-hand side. Assuming also that $k_z/\omega\sim u_d\ll \omega/(4\omega_c\gamma_p^2)$ one finally finds
 \eqb
4\pi ij_{-}^{\rm plasma}=\frac{\omega^2_p\omega}{2\omega_c^2\gamma_p^3}E_-=2\delta\omega E_-.
 \label{j_plasma}\eqe
As a consistency check, let us show that this expression for the plasma response leads to the the dispersion law (\ref{phase_velocity}).

 With this purpose, let us substitute this expression into the wave equation (\ref{eqE-}) and return to the original variable $r$. This yields
 \eqb
\frac{r_0^2}{s^{4/3}}\frac{d^2E_-}{dr^2}+2s^{2/3}\frac{r-r_0}{r_0}E_-=0.
 \eqe
In the WKB approximation, $E_-=\exp(i\int k_rdr)$, one gets
 \eqb
 k_r^2=2\frac{s^2}{r_0^3}(r-r_0).
 \label{kr}\eqe
On the other hand, the dispersion equation  (\ref{phase_velocity}) is written in the form (\ref{dispersion}), which is reduced to (\ref{kr})  at $r-r_0\ll r_0$.

 \section{Amplification factor}
 \subsection{The wave equation in the closed form}
With the results of the previous section, we can write the wave equations
in the closed form. Substituting Eqs. (\ref{rho_res}), (\ref{j_res}) and (\ref{j_plasma}) into
the equation for the "transverse" component of the electric field (\ref{eqE-}) one gets
 \eqb
 E''_-+2xE_-=\frac{\alpha}{(x-x_{\rm res}+i0)^2}
\left(1+(x-x_{\rm res})\frac{d}{dx}\right)\left(E_-+\frac{E_+}{(1-\kappa)u_d}\right);
 \label{eqE-1}\eqe
where
 \eqb
\alpha= \frac{\omega^2_bs^2u_d^6\kappa (1-\kappa)^3}{\omega^2\gamma_b};
 \label{alpha}\eqe
Taking into account that $s\approx\omega r$,  one can present $\alpha$, with the aid of Eqs. (\ref{omega_b}), (\ref{radius}) and (\ref{u_d}), as
 \eqb
\alpha=2(D/R_L)^{1/2} u_d^5\kappa (1-\kappa)^3. 
 \label{alpha1}\eqe
Taking into account that $D\lesssim R_L$, $\kappa\sim 1$ and the drift velocity is always small (recall that here $u_d$ is measured in units of $c$), one sees that $\alpha$ is a very small quantity. This already ensures that the amplification factor is small but let us calculate it explicitly.

Eq. (\ref{eqE-1}) should be generally complemented by Eqs. (\ref{eqE+}) and (\ref{eqEr}). However, Eq. (\ref{eqE-1}) is reduced to a closed equation for $E_-$ provided
 \eqb
E_+\ll (1-\kappa)u_dE_-.
 \label{neglectE+}\eqe
The conditions for such an approximation could be found as follows.

One can estimate $E_+$ from Eq. (\ref{eqEr}). Outside the resonance region, one can
write $E_+\sim s^{-1/3}E_r$ because the right-hand side of this equation is of the order of
$\delta E_-\ll s^{-1/3}E_-$. We are interested in the wave polarized predominantly in $z$ direction
therefore $E_r\ll E_-$. Then the condition (\ref{neglectE+}) is fulfilled if
\eqb
(1-\kappa)s^{1/3}u_d\gtrsim 1,
\label{cond_ud}\eqe
i.e. if the drift velocity is not too small. Taking into account that the effect under
consideration assumes non-negligible $u_d$ so that the wave amplification is suppressed
when $u_d$ goes to zero, we assume that the condition (\ref{cond_ud}) is fulfilled, which justifies the inequality (\ref{neglectE+}) outside the resonance region. It is shown in Appendix that if the inequality (\ref{neglectE+}) is fulfilled outside the resonance region, it remains valid also within the resonance region. Therefore the condition (\ref{cond_ud}) justifies neglect of $E_+$ in Eq. (\ref{eqE-1}).
Now we come to the closed equation for $E_-$
 \eqb
 E''_-+2xE_-=\frac{\alpha}{(x-x_{\rm res}+i0)^2}
\left(E_-+(x-x_{\rm res})\frac{dE_-}{dx}\right);
 \label{eqE-2}
 \eqe
Inasmuch as $\alpha\ll 1$, this equation could be solved by the method of matching asymptotic expansions (e.g. \citet{nayfe}).

\subsection{Solutions outside and inside the resonance zone}

At $|x-x_{\rm res}|\gg\sqrt{\alpha}$, one can neglect the term the right-hand side of Eq. (\ref{eqE-2}); then the solutions are expressed via the Airy functions. Specifically to the left of the resonance, at $x_{\rm res}-x\gg\sqrt{\alpha}$, we have to choose the evanescent wave solution
 \eqb
E_-^{\rm left}=\Ai(-2^{1/3}\xi). 
 \label{inner}\eqe
To the right of the resonance, at $x-x_{\rm res}\gg\sqrt{\alpha}$, the  external solution
should be chosen in the general form
 \eqb
E_-^{\rm right}=A_1\Ai(-2^{1/3}\xi)+A_2\Bi(-2^{1/3}\xi);
 \label{outer}\eqe
the amplitudes $A_1$ and $A_1$ should be found by matching the solution (\ref{outer})
to the solution (\ref{inner}) via the resonance layer.
In the wave zone, $x\gg 1$, the function (\ref{outer}) is reduced to a superposition of an ingoing and outgoing waves
 \eqb
E_-^{\rm right}=\frac 1{2^{7/6}x^{1/2}}\left[(A_1+iA_2)e^{i
\left[\frac 13(2x)^{3/2}-\frac{\pi}4\right]}+(A_1-iA_2)e^{-i\left[\frac 13(2x)^{3/2}-\frac{\pi}4\right]}\right].
 \eqe
The amplification factor is defined as the ratio of the amplitudes of the outgoing and ingoing waves
 \eqb
a=\left|\frac{A_1+iA_2}{A_1-iA_2}\right|.
 \label{a_factor}\eqe
 We will see that $|A_2|\ll |A_1|$ (at $\alpha=0$, the full solution is described be Eq. (\ref{inner}), which yields $A_1=1$; $A_2=0$; therefore at $\alpha\ll 1$, we obtain $A_2\ll A_1$). Then one can write
 \eqb
 a=1+2\Im A_2
  \label{refl}\eqe
so that the amplification factor is determined only by the imaginary part of the  amplitude $A_2$. Now let us proceed to matching the solutions $E_-^{\rm right}$ and $E_-^{\rm left}$ via the resonance zone.

First we have to find the "inner" solution, i.e. that in the resonant zone $|x-x_{\rm res}|\ll 1$. Introducing the "inner" variable
$z=(x-x_{\rm res})/\sqrt{\alpha}$, one writes Eq. (\ref{eqE-2}) as
  \eqb
 \frac{d^2E_-}{dz^2}+2\alpha(\sqrt{\alpha}z+x_{\rm res})E_-=\frac{1}{(z+i0)^2}
\left(E_-+z\frac{dE_-}{dz}\right).
 \label{eqE-3}
 \eqe
One sees that one can neglect the second term in the left-hand side; then the solution is easily found in the power-law form. Returning to the original variable, $x$, one can write
the solution in the resonant zone as
 \eqb
E_-=Q(x-x_{\rm res}+i0)^{-\alpha}+P(x-x_{\rm res}+i0)^{1+2\alpha};
\label{res_solution}\eqe
where $Q$ and $P$ are constants.  Now we have to match the solutions (\ref{inner}), (\ref{outer})
and (\ref{res_solution}). This could be done because the "outer" solutions (\ref{inner}) and (\ref{outer})
are valid at $x_{\rm res}-x\gg\sqrt{\alpha}$ and $x-x_{\rm res}\gg\sqrt{\alpha}$, correspondingly, therefore at small $\alpha$, the domains of validity of these solutions are overlapped with the domain of validity of the "inner" solution, $|x-x_{\rm res}|\ll 1$.

\subsection{Matching the solutions}
We first find the constants $Q$ and $P$ by matching the "inner" solution (\ref{res_solution}) with the solution (\ref{inner}). In the region
\eqb
\sqrt{\alpha}\ll x_{\rm res}-x\ll 1
\label{inner_matching}\eqe
the solution (\ref{res_solution}) is still valid.
Having found $Q$ and $P$ one can find amplitudes $A_1$ and $A_2$ by matching the solution
(\ref{res_solution}) with (\ref{inner}) in the region
 \eqb
\sqrt{\alpha}\ll x-x_{\rm res}\ll 1.
\label{outer_matching}\eqe
We will see that it is not necessary to perform the full matching procedure if we are interested only in the reflection coefficient, which is determined only by the imaginary part of the amplitudes.

In the region (\ref{inner_matching}) the "inner" solution (\ref{res_solution}) is reduced to
 \eqb
E_-=Q\left(1-\alpha\ln|x-x_{\rm res}|-i\pi\alpha\right)+
P(x-x_{\rm res})\left(1+2\alpha\ln|x-x_{\rm res}|+2i\pi\alpha\right).
  \label{inner1}\eqe
The "external" solution is found from Eq. (\ref{eqE-2}) as a slightly perturbed
evanescent wave. Namely the right-hand side of this equation is small in the region  (\ref{inner_matching}) therefore one can substitute there the evanescent wave (\ref{inner}) thus coming to an inhomogeneous Airy equation. The solution to this equation, which goes to (\ref{inner}) at $x\to -\infty$, is written as
 \eqb
E_-=E_-^{\rm left} +\alpha\int_{-\infty}^x\frac{G(x,x_1)}{(x_1-x_{\rm res})^2}
\left[1+(x_1-x_{\rm res})\frac{d}{dx_1}\right]E_-^{\rm left}(x_1)dx_1;
\label{inner_sol}\eqe\eqb
G(x,x_1)=2^{1/3}\pi\left[\Ai(-2^{1/3}x_1)\Bi(-2^{1/3}x)-\Bi(-2^{1/3}x_1)\Ai(-2^{1/3}x)\right].
 \eqe
At $x\to x_{\rm res}$ the first term in this solution is reduced to
the linear function
\eqb
E_-^{\rm left}(x\to x_{\rm res})=\Ai(-2^{1/3}x_{\rm res})-2^{1/3}\Ai'(-2^{1/3}x_{\rm res})(x-x_{\rm res});
\eqe
so that matching with the "inner" solution (\ref{inner1}) to within the zeroth order in $\alpha$ yields
 \eqb
Q=\Ai(-2^{1/3}x_{\rm res})+O(\alpha);\qquad
P=-2^{1/3}\Ai'(-2^{1/3}x_{\rm res})+O(\alpha).
\label{QP}\eqe
As $x\to x_{\rm res}$, the second term in (\ref{inner_sol}) gives the terms of the order of $\alpha\ln|x-x_{\rm res}|$ and $\alpha$  therefore the solution (\ref{inner_sol}) could be smoothly matched with the "inner" solution (\ref{inner1}).

We are interested only in the imaginary parts of the coefficients $Q$ and $P$, which are sufficient in order to find the imaginary part of the amplitude $A_2$. Since the solution
(\ref{inner_sol}) is real, it is matched with the solution (\ref{inner1})
if $\Im Q=\pi\alpha\Re Q$ and $\Im P=-2\pi\alpha\Re P$ to within the first order in $\alpha$. Making use of
Eq. (\ref{QP}), one gets
 \eqb
 \Im Q=-\pi\alpha\Ai(-2^{1/3}x_{\rm res});\qquad \Im P=-2^{4/3}\pi\alpha\Ai'(-2^{1/3}x_{\rm res}).
 \label{QandP}\eqe

In the region (\ref{outer_matching}) the resonance solution (\ref{res_solution}) is written as
 \eqb
E_-=Q\left[1-\alpha\ln(x-x_{\rm res})\right]+
P(x-x_{\rm res})\left[1+2\alpha\ln(x-x_{\rm res})\right];
  \label{inner2}\eqe
whereas the "external" solution is presented as
\eqb
E_-=E_-^{\rm right} -\alpha\int_x^{\infty}\frac{G(x,x_1)}{(x_1-x_{\rm res})^2}
\left[1+(x_1-x_{\rm res})\frac{d}{dx_1}\right]E_-^{\rm right}(x_1)dx_1;
\label{ext}\eqe
 At $ x\to x_{\rm res}$, one can substitute $E_-^{\rm right}$ by a linear function
 \begin{eqnarray}
E_-^{\rm right}(x)=A_1\Ai(-2^{1/3}x_{\rm res})+A_2\Bi(-2^{1/3}x_{\rm res}) \label{rightE}\\
-2^{1/3}\left[A_1\Ai'(-2^{1/3}x_{\rm res})+A_2\Bi'(-2^{1/3}x_{\rm res}\right](x-x_{\rm res}).\nonumber
 \end{eqnarray}
The second term in Eq. (\ref{ext}) has the order of $\alpha\ln(1/\alpha)$ in the region (\ref{outer_matching}) because the integral diverges only logarithmically. Therefore matching of the solution (\ref{ext}) with (\ref{inner2}) in the zeroth in $\alpha$ approximation yields, with account of
Eq. (\ref{QP}),
 \eqb
A_1=1+O(\alpha);\qquad A_2=O(\alpha).
 \label{A}\eqe
In this case, the general  expression for the amplification factor (\ref{a_factor}) is reduced to Eq. (\ref{refl}). Therefore it suffices to find only the imaginary part of $A_2$.

 With this purpose, we can only match the imaginary parts of (\ref{inner2}) and  (\ref{ext}) at $x\to x_{\rm res}$. With account of Eq. (\ref{QandP}), one gets from Eq. (\ref{inner2})
 \eqb
\Im  E_-=-\pi \alpha\left[\Ai(-2^{1/3}x_{\rm res})+2^{4/3}\Ai'(-2^{1/3}x_{\rm res})(x-x_{\rm res})\right]+O(\alpha^2).
 \eqe
Eq. (\ref{ext}) yields at $ x\to x_{\rm res}$, with account of Eqs. (\ref{rightE}) and (\ref{A}),
 \begin{eqnarray}
\Im E_-=\Im A_1\Ai(-2^{1/3}x_{\rm res})+\Im A_2\Bi(-2^{1/3}x_{\rm res}) \\
-2^{1/3}\left[\Im A_1\Ai'(-2^{1/3}x_{\rm res})+\Im A_2\Bi'(-2^{1/3}x_{\rm res}\right](x-x_{\rm res})+O(\alpha^2).\nonumber
 \end{eqnarray}
Comparing the coefficients at $ (x- x_{\rm res})^0$ and $ (x-x_{\rm res})$ in the two last equations one gets
 \begin{eqnarray}
\Im A_1 & =& -\alpha\pi^2\left[\Ai(-2^{1/3}x_{\rm res})\Bi'(-2^{1/3}x_{\rm res})+2^{4/3}\Ai'(-2^{1/3}x_{\rm res})\Bi(-2^{1/3}x_{\rm res})\right];   \\
\Im A_2 &=& \left(1+2^{4/3}\right)\alpha\pi^2\Ai(-2^{1/3}x_{\rm res})\Ai'(-2^{1/3}x_{\rm res}).
 \end{eqnarray}
Here we used the formula for Wronskian, $\Ai\Bi'-\Ai'\Bi=1/\pi$.
Now  Eq. (\ref{refl}) yields finally
\eqb
a=1+2\left(1+2^{4/3}\right)\pi^2\alpha \Ai(-2^{1/3}x_{\rm res})\Ai'(-2^{1/3}x_{\rm res}).
\label{amplification}\eqe
One sees that the wave is amplified, $a>1$, provided $\Ai(-2^{1/3}x_{\rm res})\Ai'(-2^{1/3}x_{\rm res})>0$, which happens if $x_{\rm res}$ falls within the intervals
 \eqb
 1< 2^{1/3}x_{\rm res}<2.4;\qquad 3.2<2^{1/3}x_{\rm res}<4;\qquad\dots\qquad
 \left[\frac 38(1+4k)\pi\right]^{2/3}< 2^{1/3}x_{\rm res}< \left[\frac 38(3+4k)\pi\right]^{2/3},
 \eqe
where $k$ is a large integer number.

It follows immediately from  Eq. (\ref{xres}) that  the necessary condition for the amplification,
$x_{\rm res}>2^{-1/3}$, is achieved only if $\delta$, i.e. the deviation of the wave velocity from $c$, is not too small:
 \eqb
\delta>\frac 12(1-\kappa)^2u_d^2+\frac 1{2^{1/3}s^{2/3}}.
 \label{cond_delta}\eqe
 Taking into account the condition (\ref{cond_ud}), one sees that the condition (\ref{cond_delta})
roughly coincides with the condition (\ref{cond_WKB}) found in the WKB approximation. This condition implies that the curvature drift instability is possible only if the pulsar plasma density if high enough.

In any case one sees from Eq. (\ref{amplification}) that even the instability condition is fulfilled, the amplification factor (\ref{amplification}) is very close to unity because it follows from eq. (\ref{alpha1}) that $\alpha$ remains small at any reasonable conditions. Therefore the curvature drift instability could not amplify small fluctuations up to high brightness temperatures characteristic for pulsar radio emission.

\section{Conclusions}

In this paper, we rederived the curvature-drift instability, which has long been considered as a viable mechanism for pulsar radio emission. Our approach differs from that adopted in the previous works on the topic.  Namely we did not look for the growth rate of the instability but instead explicitly calculated propagation of the electromagnetic wave through the plasma moving along the curved magnetic field lines. With such an approach, we found the amplification factor of the waves. In the standard approach, the amplification factor depends on the growth rate and also on the size of the resonance zone. In the case under consideration both the growth rate and the resonance zone are determined by the curvature radius therefore there is no sense to calculate both these quantities separately. Moreover, in the previous works the growth rate was found by the dielectric tensor method,  which implies that
the waves are considered as locally plane. Then one could define the dielectric tensor of the medium such that the growth rate is an imaginary part of an appropriate eigenvalue. Even in the most advanced work by \citet{lyut_mach_bland99,lyut_bland_mach99}, where it was explicitly demonstrated that the resonance occurs in the vicinity of the caustic zone so that the wave should be described by the Airy type functions, the amplification was calculated only at the WKB condition $x_{\rm res}\gg 1$.
Our approach is free from these limitation and moreover provides the amplification factor straightforwardly without intermediate steps.

The advantage of explicit solutions over the dielectric tensor approach was clearly demonstrated by the long-lasting discussion initiated by \citet{beskin_etal87a,beskin_etal87b} works.
These authors derived the dielectric tensor of a weakly inhomogeneous plasma in the infinitely strong magnetic field
and found unstable modes. The correctness of the obtained dielectric tensor has being extensively debated, see \citet{nambu89,machabeli91,machabeli95,istomin94,nambu96,bornatici_kravtsov00}.
However, the claim of Beskin et al was disproved by \citet{larroche_pellat87} who presented an explicit solution
to the wave equations and demonstrated that the wave amplitude does not grow unless the flow is sharply bounded
(see also discussion in \citet{beskin_etal88,larroche_pellat88}).
The specific property of the obtained solution is that it becomes singular in a narrow resonance layer. The dielectric tensor method assumes implicitly that the global solution could be described by a locally plane wave with smoothly varying parameters. In shear flows, there are typically no such solutions satisfying reasonable boundary conditions; the wave
becomes singular in the critical layer where the phase velocity becomes equal to the flow velocity
(e.g. \citet{stepanyants_fabrikant89}). Therefore the dielectric tensor method does not provide unambiguous conclusions on the stability of shear flows; one has to find explicit solutions as it was done in the present paper.

One more novelty of the present research is taking into account of the inertial forces (Coriolis and centrifugal) when considering interaction of the waves with resonant particles. These terms were ignored in all previous works on the curvature drift instability. We have shown that these forces significantly suppress the growth rate.

The net result of our study is that at no condition the amplification factor could become large therefore the curvature drift instability should be excluded from the list of potential mechanisms for pulsar radio emission.

\acknowledgements
This work was supported by German-Israeli Foundation for Scientific Research and Development under the grant I-804-218.7/2003 and by the Israeli Science Foundation under the grant 737/07.

\appendix
\section{Appendix. General solution to the wave equations in the resonance region.}

Within the resonance region, $|x-x_{\rm res}|\ll 1$, one has to retain only
the resonant terms so that the equations (\ref{eqE-}),
(\ref{eqE+}) and (\ref{eqEr}) could be written, with account of (\ref{rho_res}) and (\ref{j_res}), in the form
  \begin{eqnarray}
 E_-'' &=& \frac{\alpha}{(x-x_{\rm res}+i0)^2}
\left(1+(x-x_{\rm res})\frac{d}{dx}\right)\left(E_-+\frac{E_+}{(1-\kappa)u_d}\right);\label{E-''}\\
 is^{1/3}E'_+&=&2x_{\rm res}E_r;\\
(1-\kappa)u_ds^{1/3} E'_r &=& i\frac{\alpha}{(x-x_{\rm res}+i0)^2}
\left(1+(x-x_{\rm res})\frac{d}{dx}\right)\left(E_-+\frac{E_+}{(1-\kappa)u_d}\right).
 \end{eqnarray}
Eliminating $E_r$ from the last two equations and making use of Eq. (\ref{xres}), one gets
 \eqb
 E_+'' =-\frac{(1-\kappa)u_d\alpha}{(x-x_{\rm res}+i0)^2}
\left(1+(x-x_{\rm res})\frac{d}{dx}\right)\left(E_-+\frac{E_+}{(1-\kappa)u_d}\right);
 \label{E+''}\eqe
thus reducing the system to two equations for $E_-$ and $E_+$.

It follows immediately from Eqs. (\ref{E-''}) and (\ref{E+''}) that $E''_++(1-\kappa)u_dE''_-=0$, which implies
 \eqb
E_++(1-\kappa)u_dE_-= C_1+C_2(x-x_{\rm res});
 \eqe
where $C_1$ and $C_2$ are constants. Now one can find the fields in the resonance region
in the form
 \begin{eqnarray}
E_- &=& Q+P(x-x_{\rm res})+\frac{\alpha[S(x-x_{\rm res})+T]}{(1-\kappa)u_d} \ln(x-x_{\rm res}+i0);\label{E-sol}\\
E_+ &=& -T-(1-\kappa)u_dQ+\frac 12\left[S-2(1-\kappa) u_dP\right](x-x_{\rm res})
 \nonumber\\
&&-\alpha [S(x-x_{\rm res})+T]\ln(x-x_{\rm res}+i0);\label{E+sol}
 \end{eqnarray}
where $P$, $Q$, $S$ and $T$ are constants. These constants could be found by matching, at $|x-x_{\rm res}|\sim 1$, with the solution outside of the resonance region.

It has been shown in sect. 4 that at the condition (\ref{cond_ud}), the "longitudinal"
component, $E_+$, satisfies the inequality (\ref{neglectE+}) outside the resonance region.
Inspecting Eqs. (\ref{E-sol}) and (\ref{E+sol}) one sees that this inequality is fulfilled at $|x-x_{\rm res}|\sim 1$
provided
 \eqb
\vert T+(1-\kappa)u_dQ\vert\ll (1-\kappa)u_dQ;
 \eqe
  \eqb
\left\vert S-2(1-\kappa)u_dP\right\vert\ll (1-\kappa)u_dP;
 \eqe
which yields
\eqb
 T=-(1-\kappa)u_dQ;\qquad S=2(1-\kappa)u_dP.
 \label{TS}\eqe
One immediately sees that in this case the inequality (\ref{neglectE+}) is satisfied also within the resonance region with the exception of the negligibly small region $|x-x_{\rm res}|\lesssim\exp(-1/\alpha)$.
Therefore one can neglect $E_+$ in the resonance term (that in the right-hand side of Eq.
(\ref{eqE-1})) thus coming to the closed equation (\ref{eqE-2}). This justifies use of
Eq. (\ref{eqE-2}) in order to find the reflection coefficient. Note that upon substituting Eq. (\ref{TS}) into the general resonance solution (\ref{E-sol}) one comes to  Eq. (\ref{inner2}) obtained as the solution to Eq. (\ref{eqE-2}).


\bibliographystyle{apj}
\bibliography{pulsar}

\end{document}